\documentclass[preprint,12pt]{elsarticle}

\usepackage{amssymb}
\usepackage{amsmath}
\usepackage{upgreek}
\usepackage{booktabs}
\usepackage{multirow}
\usepackage{multicol}
\usepackage{ulem}
\usepackage{xcolor}
\newcommand{\apjl}{The Astrophysical Journal Letters}

\newcommand{\aap}{Astron. Astrophys.}
\newcommand{\pasp}{Publ. Astron. Soc. Pac}
\newcommand{\pasj}{Publ. Astron. Soc. Jpn}
\newcommand{\ssr}{Space Science Reviews}
\newcommand{\planss}{Planetary and Space Science}

\journal{Nuclear Inst. and Methods in Physics Research, A}

\begin{document}

\begin{frontmatter}



\title{Long-term stability of scientific X-ray CMOS detectors}

\author[a,b]{Mingjun Liu}
\author[a,b]{Qinyu Wu}
\author[a,b,c]{Zhixing Ling\corref{cor1}}
\cortext[cor1]{Corresponding authors.}
\ead{lingzhixing@nao.cas.cn}

\author[a,b]{Chen Zhang}
\author[a,b]{Weimin Yuan}
\author[a,b,d]{Shuang-Nan Zhang}

\affiliation[a]{organization={National Astronomical Observatories, Chinese Academy of Sciences},
             city={Beijing},
             postcode={100101},
             country={People's Republic of China}}

\affiliation[b]{organization={School of Astronomy and Space Science, University of Chinese Academy of Sciences},
             city={Beijing},
             postcode={100049},
             country={People's Republic of China}}

\affiliation[c]{organization={Institute for Frontiers in Astronomy and Astrophysics, Beijing Normal University},
             city={Beijing},
             postcode={102206},
             country={People's Republic of China}}

\affiliation[d]{organization={Institute of High Energy Physics, Chinese Academy of Sciences},
            city={Beijing},
            postcode={100049}, 
            country={People's Republic of China}}

\begin{abstract}
In recent years, complementary metal-oxide-semiconductor (CMOS) sensors have been demonstrated to have significant potential in X-ray astronomy, where long-term reliability is crucial for space X-ray telescopes. This study examines the long-term stability of a scientific CMOS sensor, focusing on its bias, dark current, readout noise, and X-ray spectral performance. The sensor was initially tested at -30 $^\circ$C for 16 months, followed by accelerated aging at 20 $^\circ$C. After a total aging period of 610 days, the bias map, dark current, readout noise, gain, and energy resolution exhibited no observable degradation. There are less than 50 pixels within the 4 k $\times$ 4 k array which show a decrease of the bias under 50 ms integration time by over 10 digital numbers (DNs). First-order kinetic fitting of the gain evolution predicts a gain degeneration of 0.73\% over 3 years and 2.41\% over 10 years. These results underscore the long-term reliability of CMOS sensors for application in space missions.
\end{abstract}



\begin{keyword}



X-ray detector \sep CMOS \sep aging \sep reliability

\end{keyword}

\end{frontmatter}



\section{Introduction}
\label{sec:intro}

Silicon image sensors, including charge-coupled devices (CCDs) and complementary metal-oxide-semiconductor (CMOS) detectors, dominate digital imaging applications across scientific and industrial sectors \cite{2003ARNPS..53..263J}. Due to their high quantum efficiency and low noise \cite{2017JATIS...3c6002N}, CCDs have emerged as the predominant detector type in optical and X-ray astronomy. Since the 1990s, CCDs have been extensively deployed in X-ray telescopes, including notable missions such as ASCA \cite{1993SPIE.2006..272B}, XMM-Newton \cite{2001A&A...365L..18S}, Chandra \cite{2003SPIE.4851...28G}, Swift \cite{2005SSRv..120..165B}, Suzaku \cite{2007PASJ...59S...1M}, MAXI \cite{2009PASJ...61..999M}, eROSITA \cite{2021A&A...647A...1P}, and XRISM \cite{2024SPIE13093E..1IM}. In these missions, CCDs have demonstrated exceptional performance in terms of position resolution, energy resolution, and reliability. In contrast, CMOS sensors, characterized by their low cost, reduced energy consumption, fast readout speeds, and higher working temperatures, have recently found their way into astronomical applications \cite{2009SPIE.7419E..0TH,2022JATIS...8b6001R}. The newly launched X-ray missions, Lobster Eye Imager for Astronomy (LEIA)\cite{2022ApJ...941L...2Z,2023RAA....23i5007L} and Einstein Probe (EP) \cite{2018SPIE10699E..25Y, 2022hxga.book...86Y}, have substantially used CMOS sensors for the first time. Upcoming missions like THESEUS \cite{2020SPIE11454E..0IH} and JUICE \cite{2013P&SS...78....1G}, also plan to utilize CMOS detectors to achieve their scientific goals.

Space telescopes face harsh challenges regarding long-term reliability in orbit compared to ground-based devices. Although on-orbit service (OOS) was proposed in the 1960s, only a few satellites have received OOS \cite{2019PrAeS.108...32L}, such as the Hubble Space Telescope (HST) and the Solar Maximum Satellite (SMM). Many space projects, especially those aimed at deep space exploration, must operate for several years without any OOS. Therefore, understanding the long-term stability of detectors is crucial for the success of space missions. To our knowledge, studies investigating the stability of CCDs and CMOS sensors are quite limited \cite{WOS:000264367300014,WOS:000073452500015,WOS:000426442400001}, with research on the long-term aging of CMOS detectors on an annual scale being especially rare. In addition, the long-term aging effects cannot be decoupled from the radiation damages using the in-orbit data of space telescopes with the ubiquitous high-radiation environment.

In our previous work on scientific CMOS sensors, we have investigated the effects of structure design \cite{2018SPIE10699E..5OW,2022JInst..17P2006H,2022JInst..17P2016W}, developed a set of methods for obtaining high-quality X-ray spectra \cite{2021JInst..16P3018L,2022SPIE12191E..0LW,2023PASP..135b5003W}, and demonstrated the excellent X-ray performances \cite{2019JInst..14P2025W,2022PASP..134c5006W,2023NIMPA105068180W,2023PASP..135k5002W} and high radiation tolerance \cite{2023JATIS...9d6003L,10.1117/1.JATIS.10.2.026001} of scientific CMOS sensors for X-ray astronomical observations. In this work, we study the long-term stability of a customized large-format scientific CMOS sensor, named EP4K \cite{2023PASP..135k5002W}, which has been extensively deployed in the LEIA and EP missions that are currently in orbit. The experimental setup and data processing methods are described in Section ~\ref{sec:exp}. The stability of bias, dark current, noise, and X-ray spectrum are presented in Section ~\ref{sec:res}. The main conclusions are summarized and discussed in Section ~\ref{sec:con}.

\section{Experimental Setup}\label{sec:exp}

The aging experiment was carried out at National Astronomical Observatories, Chinese Academy of Sciences from February 2022 to October 2023. One EP4K scientific CMOS sensor, which is from the same batch as those installed on the Wide-field X-ray Telescope (WXT) on board the EP satellite, was employed in the aging experiment. EP4K has a 4 k $\times$ 4 k pixel array with a pixel size of 15 $\upmu$m $\times$ 15 $\upmu$m and has an epitaxial layer of 10 $\upmu$m (see \cite{2022PASP..134c5006W,2023PASP..135k5002W} for detailed performances). Throughout the aging process, this sensor was initially aged at -30 $^\circ$C until June 2023, followed by a temperature adjustment to 20 $^\circ$C until the end of the experiment. The aging experiment was conducted in a vacuum environment below 0.1 Pa. A camera developed by our laboratory \cite{2022SPIE12191E..0LW} was used to operate the CMOS sensor at its current maximum frame rate of 20 Hz, allowing for real-time temperature control and data readout.

As shown in Table \ref{tab:test}, every a few months, a series of dark exposure tests and X-ray exposure tests of the sensor were carried out. At various temperatures, mainly spanning from -30 $^\circ$C to 20 $^\circ$C, we recorded the basic properties of the scientific CMOS sensor. These included dark current, readout noise, and bias, which were extracted from dark exposures, while conversion gain and energy resolution were derived from exposure to an $^{55}$Fe source. All data were collected under the high gain mode, with the programmable gain amplifier (PGA) register set to 7.5 (see \cite{2022PASP..134c5006W} for more details). The data extraction and event grading processes followed the standard procedures given in \cite{2021JInst..16P3018L,2022PASP..134c5006W}.  

\begin{table}[ht]
    \caption{The dates and aging-temperature on which dark exposure tests and X-ray exposure tests were conducted during the aging experiment.}
    \label{tab:test}
    \begin{center}
    \begin{tabular}[width=\columnwidth]{ccccc}
    \hline\hline
    \rule[-1ex]{0pt}{3.5ex} \multirow{2}{*}[-0.5ex]{Date} & Time & Aging-temperature & \multicolumn{2}{c}{Test items} \\
    \cmidrule(r){4-5} 
    \rule[-1ex]{0pt}{3.5ex} & (d) & ($^\circ$C) & Dark exposure & X-ray exposure\\
    \hline\hline
    \rule[-1ex]{0pt}{3.5ex} 2022-02-14 & 0 & -30 & $\checkmark$ & $\checkmark$ \\ 
    \rule[-1ex]{0pt}{3.5ex} 2022-03-18 & 32 & -30 & $\checkmark$ & $\checkmark$ \\
    \rule[-1ex]{0pt}{3.5ex} 2022-05-19 & 94 & -30 & $\checkmark$ & $\checkmark$ \\
    \rule[-1ex]{0pt}{3.5ex} 2022-06-28 & 134 & -30 & $\checkmark$ & $\checkmark$ \\
    \rule[-1ex]{0pt}{3.5ex} 2022-08-31 & 198 & -30 & $\checkmark$ & $\checkmark$ \\
    \rule[-1ex]{0pt}{3.5ex} 2022-11-01 & 260 & -30 & $\checkmark$ & $\checkmark$ \\
    \rule[-1ex]{0pt}{3.5ex} 2023-01-03 & 323 & -30 & $\checkmark$ & $\checkmark$ \\
    \rule[-1ex]{0pt}{3.5ex} 2023-04-17 & 427 & -30 & $\checkmark$ & $\checkmark$ \\
    \rule[-1ex]{0pt}{3.5ex} 2023-08-10 & 542 & 20 & $\checkmark$ & $\times$ \\
    \rule[-1ex]{0pt}{3.5ex} 2023-09-21 & 584 & 20 & $\times$ & $\checkmark$ \\
    \rule[-1ex]{0pt}{3.5ex} 2023-10-17 & 610 & 20 & $\checkmark$ & $\checkmark$ \\
    \hline\hline
    \end{tabular}
    \end{center}
\end{table}

We primarily focus on the sensor performance under operating conditions identical to those of the EP-WXT, specifically a readout frame rate of 20 Hz (50 ms integration time) at -30 $^\circ$C. X-ray spectrum tests at 20 $^\circ$C have been included since April 2023. 

\begin{figure}
\centering
\includegraphics[width=\columnwidth]{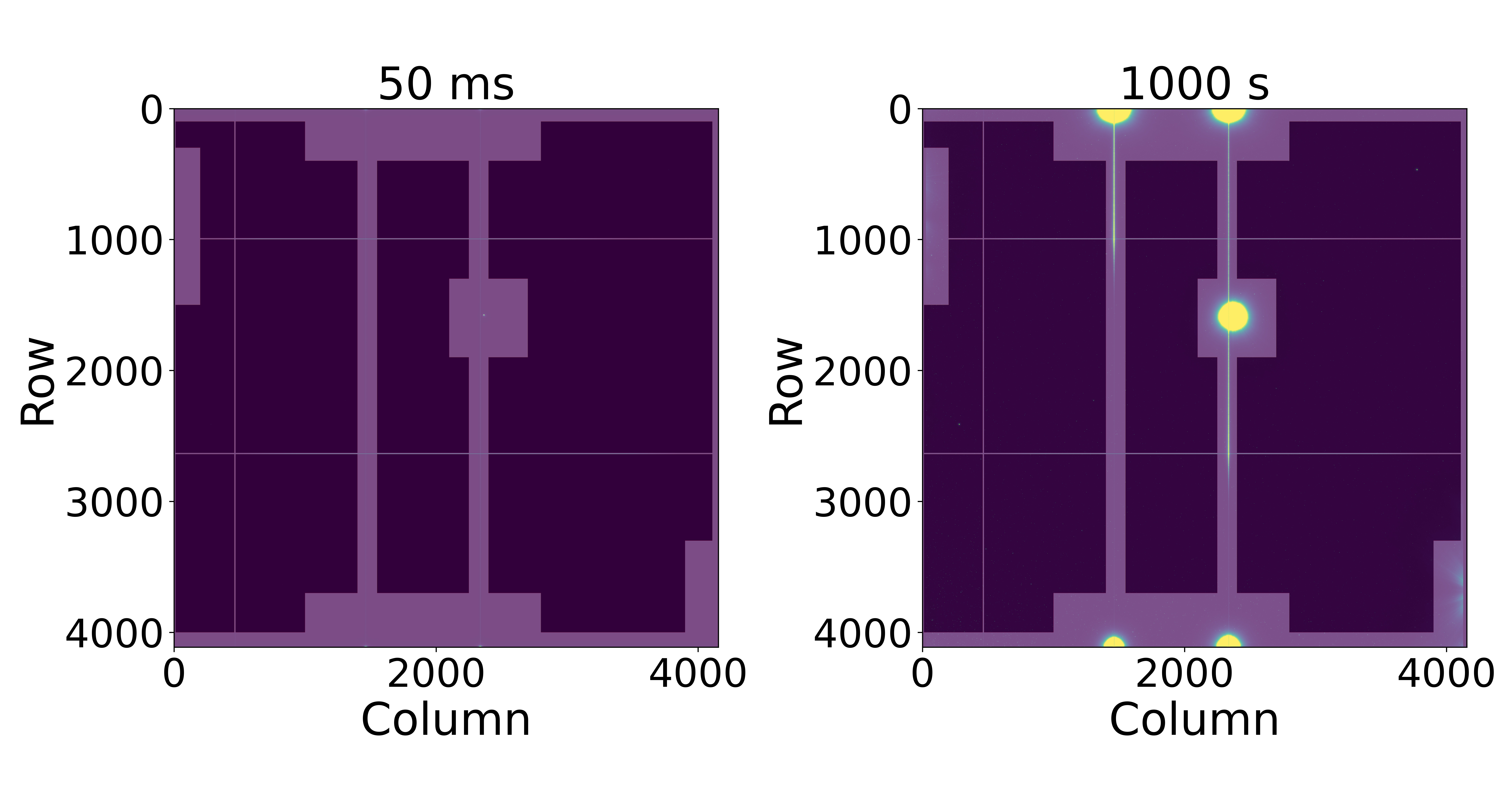}
\caption{The dark exposure map under 50 ms (left) and 1 ks (right) integration time of the sensor on the 610th day, with the defective areas (yellow) and masked regions (light purple) shown.}\label{fig:mask}
\end{figure}

We have omitted some areas (around 3.9$\times$10$^6$ pixels, i.e., 23\% of total pixels) with inherent defects or light leaks in our data analysis so that we can appropriately exhibit the long-term stability of this scientific CMOS sensor. Figure \ref{fig:mask} shows the regions excluded from data analysis.

\section{Performances during Aging}\label{sec:res}

\subsection{Bias map}\label{subsec:bias}

In CMOS sensors, each pixel has its own bias level. To evaluate the bias level of the whole sensor, we calculated the median digital number (DN) for each pixel among a set of images at single integration time to assemble the bias map. The numbers of dark frames are $\sim$50, $\sim$50, $\sim$40, 7, 3 and 2 for the integration times of 14 $\upmu$s, 50 ms, 1 s, 10 s, 100 s and 1000s, respectively. The integration times of $\sim$14 $\upmu$s and 50 ms were selected for this purpose. The 14 $\upmu$s signal can be seen as the the true bias level for the undetectable contributions from dark current at 14 $\upmu$s integration time (see Section \ref{subsec:dc} for detail). It should be noted that we use the convenient term '50-ms-bias' to refer to the signal under 50 ms integration time for two reasons. Firstly, these signals are close to the bias values under 14 $\upmu$s integration time because of the low dark current level. Secondly, the EP-WXT operated at a readout frame rate of 20 Hz.

\begin{figure}
    \centering
    \includegraphics[width=\columnwidth]{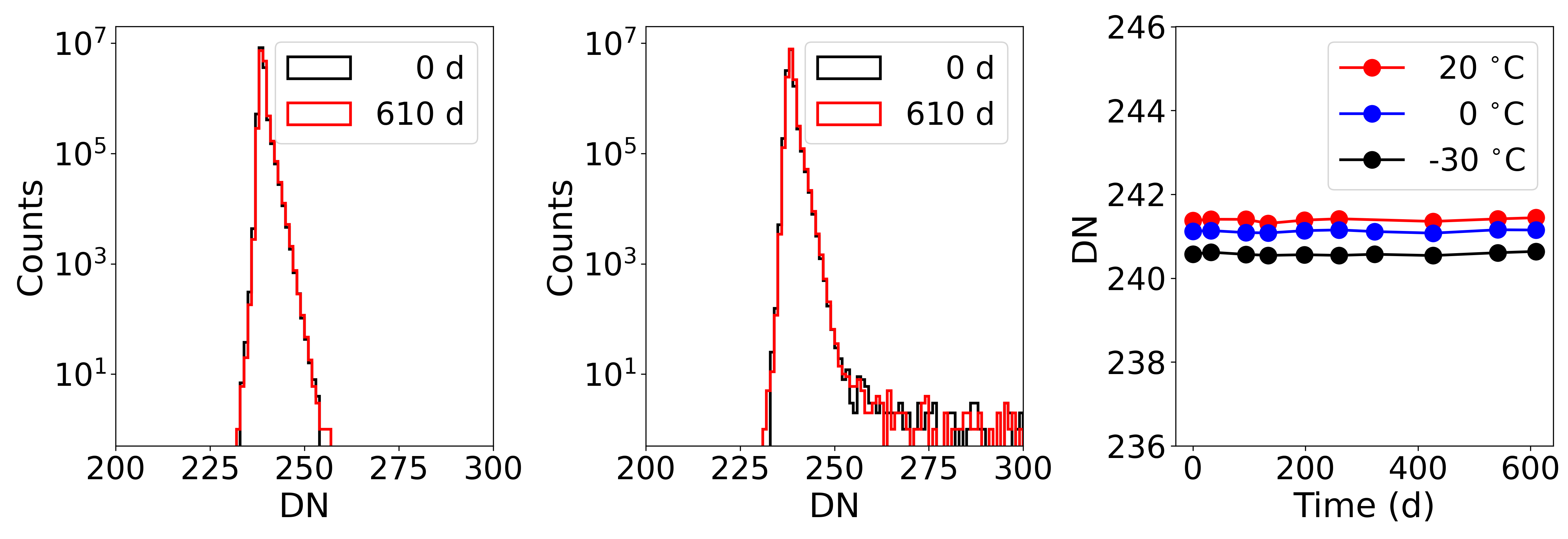}
    \caption{The distribution of bias under 14 $\upmu$s (left panel) and 50 ms (middle panel) integration time of the scientific CMOS sensor at -30 $^\circ$C on the 0th day (black) and the 610th day (red) of the aging test. It should be noted that there are 57 pixels and 54 pixels that have a 50-ms-bias value over 300 DN on the 0th day and the 610th day, respectively. The bias of the scientific CMOS sensors under 14 $\upmu$s integration time (right panel) during the aging test at -30 $^\circ$C (black), 0 $^\circ$C (blue) and 20 $^\circ$C (red).}
    \label{fig:bias}
\end{figure}

The bias distribution under the 14 $\upmu$s integration time is unchanged after the aging of 610 days as shown in the left panel of Figure \ref{fig:bias}. The median value of the bias map is used to evaluate the bias level of the whole sensor. From the right panel of Figure \ref{fig:bias}, the median value of the 14-$\upmu$s bias map can remain unchanged for over 600 days under any temperature condition ranging from -30 $^\circ$C to 20 $^\circ$C. The 50-ms bias map also shows no degeneration. 

\subsection{Dark current}\label{subsec:dc}

\begin{figure}
    \centering
    \includegraphics[width=\columnwidth]{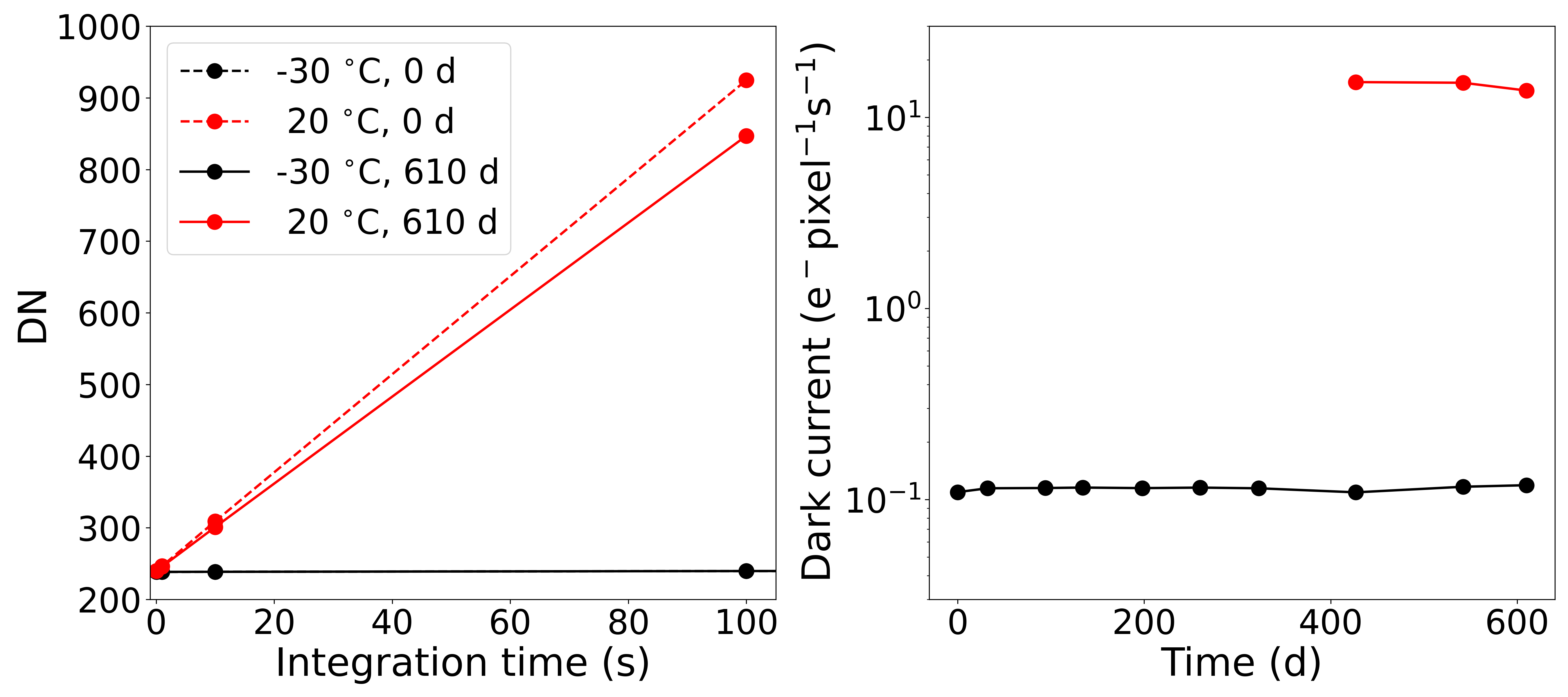}
    \caption{The bias level of the whole sensor as a function of the integration time (left panel). The dark currents of the scientific CMOS sensor at -30 $^\circ$C (black) and 20 $^\circ$C (red) during the aging test (right panel).}
    \label{fig:dc}
\end{figure}

Dark current is a vital factor for silicon image sensors, which represents the self-generated signal level without incident photons. The contribution from dark current to bias level increases almost linearly with integration time as shown in the left panel of Figure \ref{fig:dc}. Thus, we evaluate the dark current level using the slope, i.e., the bias increasing rate, which can be obtained from the mean DN values of the whole sensor under the shortest (14 $\upmu$s) integration time and the longest integration time before over-exposure (100 s or 1000s under different temperatures). As shown in right panel of Figure \ref{fig:dc}, the dark currents of this scientific CMOS sensor at -30 $^\circ$C and 20 $^\circ$C show no obvious degeneration during the whole aging period. 

\subsection{Noise}\label{subsec:std}

\begin{figure}
    \centering
    \includegraphics[width=\columnwidth]{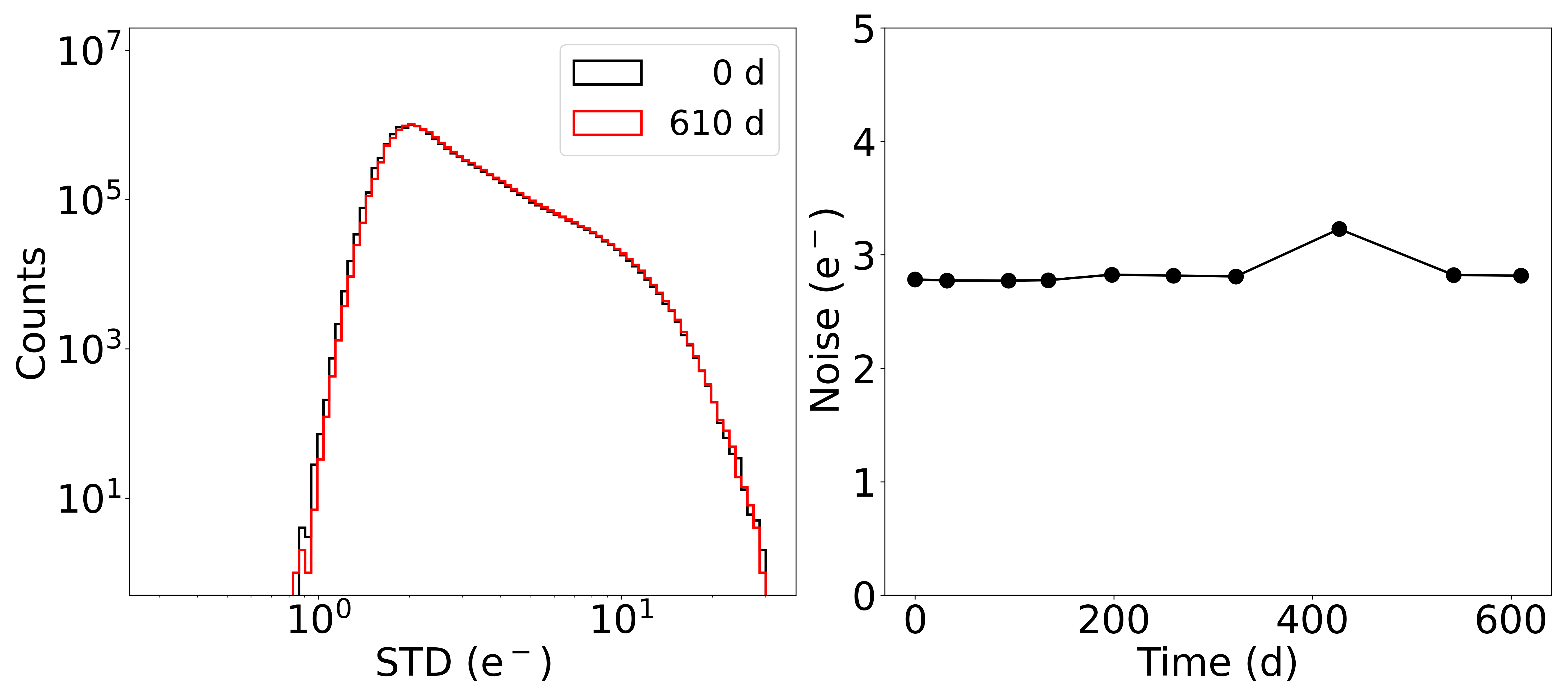}
    \caption{The distribution of the readout noise of pixels of the scientific CMOS sensor under 14 $\upmu$s integration time at -30 $^\circ$C, on the 0th day (black) and the 610th day (red) of the aging test (left panel). The right panel gives the readout noise of the sensor at -30 $^\circ$C during the aging, where the variation in April 2023 (427th day) may originate from some issues in the experimental setup.}
    \label{fig:std}
\end{figure}

The readout noise affects the quality of images. It is produced during the signal readout process in pixels relating to the readout circuit. The readout noise of each pixel is represented by the standard deviation (STD) of bias value under the minimum integration time of 14 $\upmu$s. The median of the STD distribution among pixels is used to represent the readout noise level of the whole sensor. From Figure \ref{fig:std}, the STD distribution and the readout noise of the sensor exhibit negligible degradation, except in April 2023, which may have been affected by some setup issues.

\subsection{X-ray spectrum}\label{subsec:gain}

The performance on the X-ray spectrum for CMOS sensors is described by the conversion gain (hereafter gain) and energy resolution. The gain can be understood as the conversion factor of the incident photon energy to the recorded digital signal, while the energy resolution is evaluated by the full widths at half maximum (FWHM) of emission lines.

Under a 50 ms integration time and high gain mode (settings identical to those on the EP satellite), an $^{55}$Fe source is used for the X-ray exposure experiments. Following the standard data extraction and event grading procedures described in \cite{2022PASP..134c5006W}, two types of spectra are built: GAtotal and G0center spectrum. The GAtotal spectrum is extracted from events of all kinds of grades, and the total charge in the 3 $\times$ 3 region of each event is used. In contrast, the G0center spectrum is built from the center pixel’s charge of each single pixel event or Grade 0 event.  

For each type of spectrum, we then obtain the gain by linearly fitting the positions of the centroid of emission lines, i.e., $y=a_1x+a_0$ where $x$ is the location of emission lines in the unit of DN, $y$ is the energy of the emission lines and $a_1$ is the gain. The FWHM at a given peak is obtained through the Gaussian fitting of the peak.

\begin{figure}
\centering
\includegraphics[width=0.8\columnwidth]{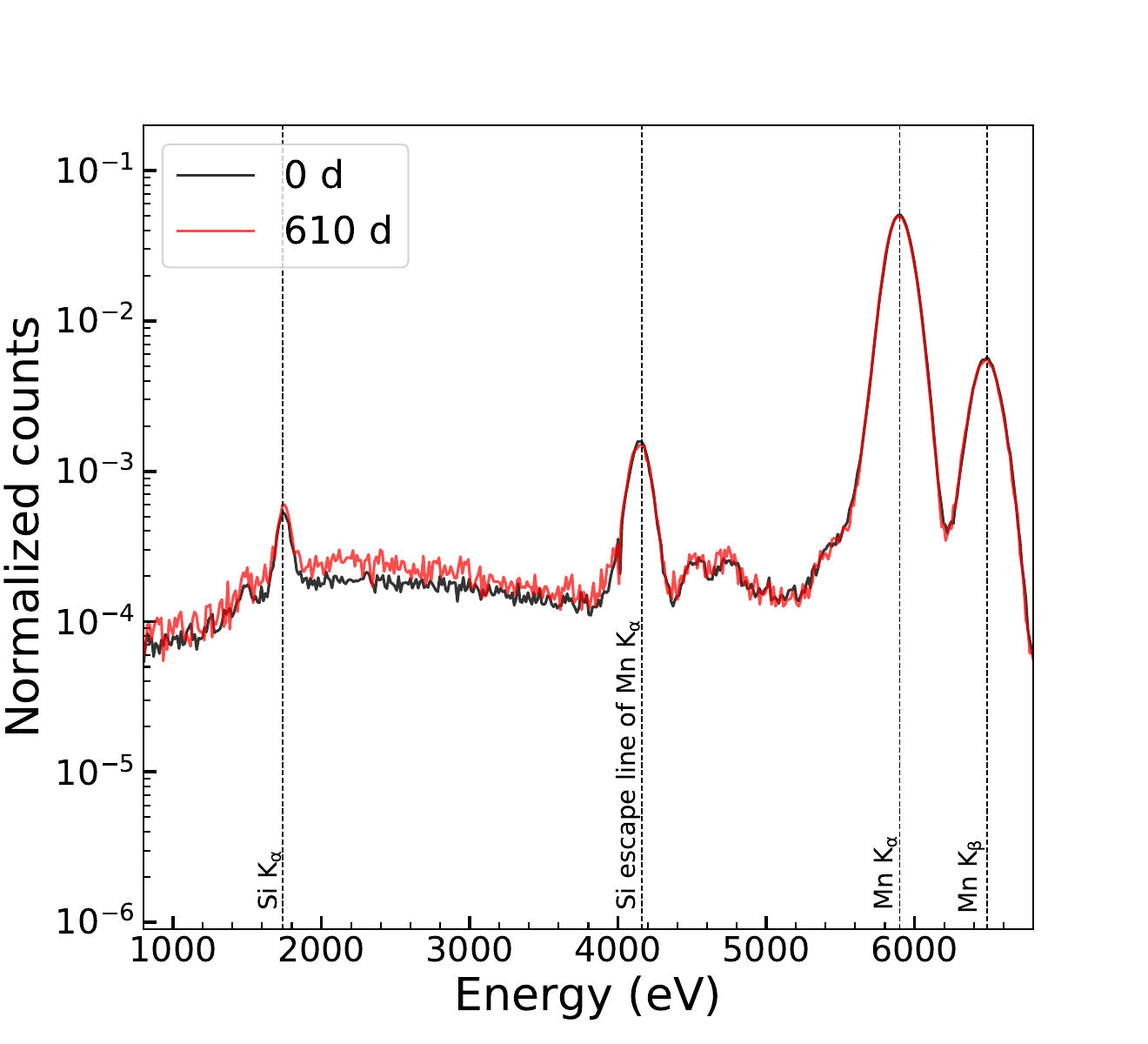}
\caption{The G0center spectra of the same sensor irradiated by a $^{55}\!\rm{Fe}$ source on the 0th day (black) and the 610th day (red) of the aging test, normalized based on the total photon counts. Four characteristic lines can be identified: Si $\rm{K_\alpha}$ (1.74 keV), Si escape peak of Mn $\rm{K_\alpha}$ (4.16 keV), Mn $\rm{K_\alpha}$ (5.90 keV), and Mn $\rm{K_\beta}$ (6.49 keV). The slight difference in the low-energy continuum is primarily attributable to the decay of the $^{55}\!\rm{Fe}$ source rather than changes in detector performance.}
\label{fig:spec}
\end{figure}

As shown in Table \ref{tab:gain}, for G0center spectra, the gains and intercepts $a_0$ remain unchanged at both -30 $^\circ$C and 20 $^\circ$C during the 610-day aging test. The energy resolutions exhibit no significant degeneration. The FWHM of the Mn K$_\alpha$ line (at 5.9 keV) only slightly varies within a few eV. The G0center spectrum is almost unchanged after the aging test as shown in Figure \ref{fig:spec}. The slight difference in the low-energy continuum is primarily attributable to the decay of the $^{55}\!\rm{Fe}$ source rather than changes in detector performance. The variation of GAtotal spectra is similar to that of G0center spectra and is also quite subtle.

\begin{table}[ht]
    \caption{The gains, intercepts and energy resolutions of G0center spectra during the aging test.}
    \label{tab:gain}
    \begin{center}
    \begin{tabular}[width=\columnwidth]{ccccccccc}
    \hline\hline
    \rule[-1ex]{0pt}{3.5ex} \multirow{2}{*}[-1ex]{Time (d)} & \multicolumn{2}{c}{Gain (eV/DN)} & \multicolumn{2}{c}{Intercept (eV)} & \multicolumn{2}{c}{FWHM at Mn K$_\alpha$ (eV)} \\ \cmidrule(r){2-3} \cmidrule(r){4-5} \cmidrule(r){6-7}
    \rule[-1ex]{0pt}{3.5ex} & -30$^{\circ}$C & 20$^{\circ}$C & -30$^{\circ}$C  & 20$^{\circ}$C  & -30$^{\circ}$C & 20$^{\circ}$C\\
    \hline\hline
    \rule[-1ex]{0pt}{3.5ex} 0 & 6.62$\pm$0.02 & / & 41$\pm$17 & / & 197.8$\pm$1.1 & / \\ 
    \rule[-1ex]{0pt}{3.5ex} 32 & 6.62$\pm$0.02 & / & 40$\pm$17  & / & 197.7$\pm$1.1 & / \\
    \rule[-1ex]{0pt}{3.5ex} 94 & 6.62$\pm$0.02 & / & 35$\pm$22  & / & 198.6$\pm$1.3 & / \\
    \rule[-1ex]{0pt}{3.5ex} 134 & 6.62$\pm$0.02 & / & 30$\pm$17  & / & 194.2$\pm$1.6 & / \\
    \rule[-1ex]{0pt}{3.5ex} 198 & 6.62$\pm$0.02 & / & 37$\pm$18 & / & 197.4$\pm$1.2 & / \\
    \rule[-1ex]{0pt}{3.5ex} 260 & 6.61$\pm$0.02 & / & 40$\pm$19  & / & 197.4$\pm$1.1 & / \\
    \rule[-1ex]{0pt}{3.5ex} 323 & 6.61$\pm$0.02 & / & 50$\pm$20 & /  & 198.2$\pm$1.1 & / \\
    \rule[-1ex]{0pt}{3.5ex} 427 & 6.60$\pm$0.02 & 6.49$\pm$0.03 & 36$\pm$14 & 31$\pm$24  & 194.1$\pm$0.9 & 185.6$\pm$1.0 \\
    \rule[-1ex]{0pt}{3.5ex} 584 & 6.61$\pm$0.02 & 6.50$\pm$0.02 & 37$\pm$20  & 24$\pm$19 & 197.9$\pm$1.2 & 190.2$\pm$1.1 \\
    \rule[-1ex]{0pt}{3.5ex} 610 & 6.61$\pm$0.03 & 6.50$\pm$0.03 & 41$\pm$25 & 17$\pm$24 & 199.7$\pm$1.3 & 189.5$\pm$1.2 \\
    \hline\hline
    \end{tabular}
    \end{center}
\end{table}

\subsection{Degraded Pixel}\label{subsec:pixel}

\begin{figure*}
    \centering
    \includegraphics[width=\columnwidth]{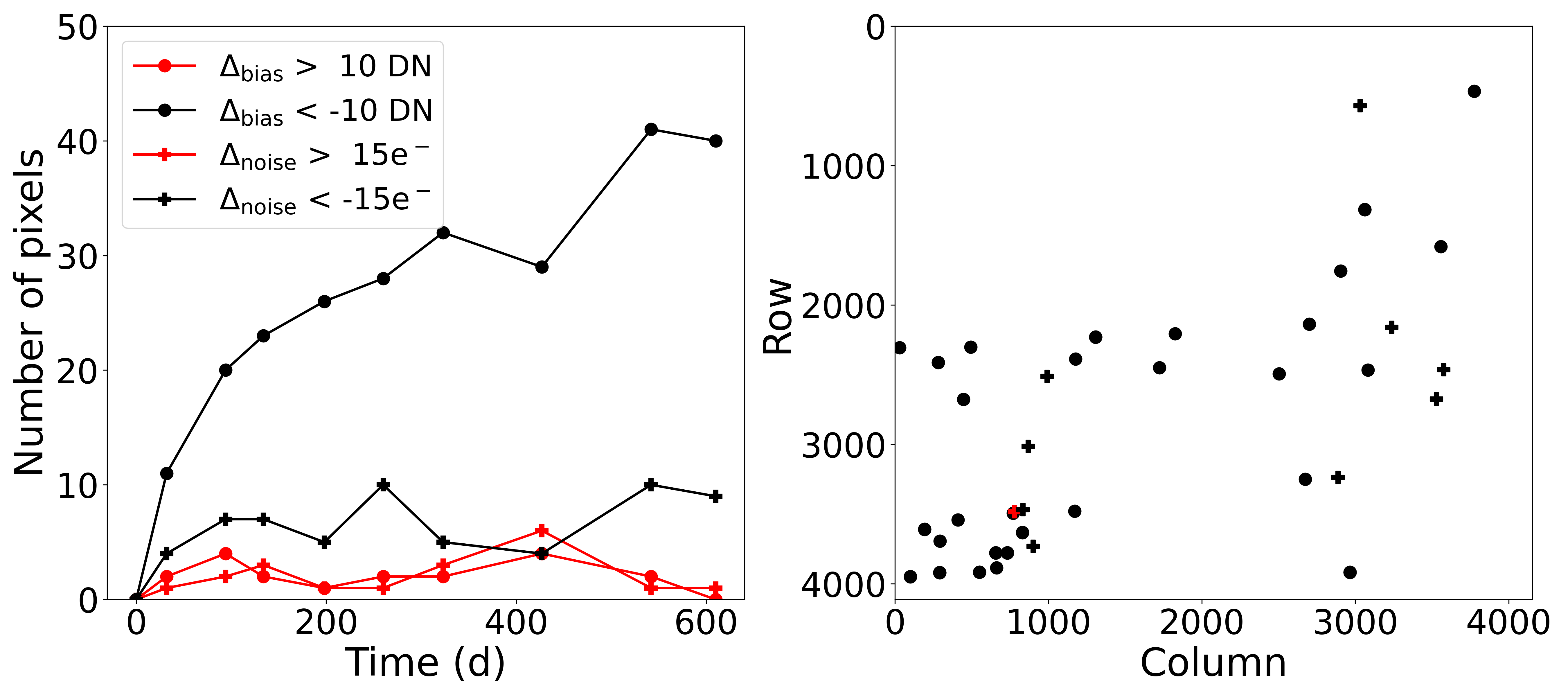}
    \caption{Left: the number of degraded pixels during the aging test. Right: the distribution of the degraded pixels on the scientific CMOS sensor at -30 $^\circ$C, after the aging test of 610 days. The points represent the number of pixels with a bias increase (red points) or decrease (black points) over 10 DN. The "+" symbols represent the number of pixels with a noise increase (red "+") or decrease (black "+") of over 15 e$^-$. }
    \label{fig:point}
\end{figure*}

High sensitivity and high-resolution astronomy observations have strict requirements on the reliability of each pixel of each detector. The basic performances of the whole sensor are essentially unchanged during the long-term aging, while a few pixels show changes in bias, dark current, and readout noise. We select degraded pixels as those whose 50-ms-bias changes over 10 DN (called bias-varied pixels) or those whose noise varies over 15 e$^-$ (called noise-varied pixels) during the aging test. As the left panel of Figure \ref{fig:point} shows, the pixels with large noise variation or large bias increment are very rare, while the number of pixels with obvious decreasing bias gradually increases during the whole aging test. On the 610th day, 40 pixels exhibit a decrease of the 50-ms-bias by over 10 DN, one pixel has a noise increment of over 15 e$^-$, and 9 pixels show a decrease of noise by over 15 e$^-$. Their distribution on the scientific CMOS sensor is non-clustering, as shown in the right panel of Figure \ref{fig:point}.

\begin{figure*}
    \centering
    \includegraphics[width=\columnwidth]{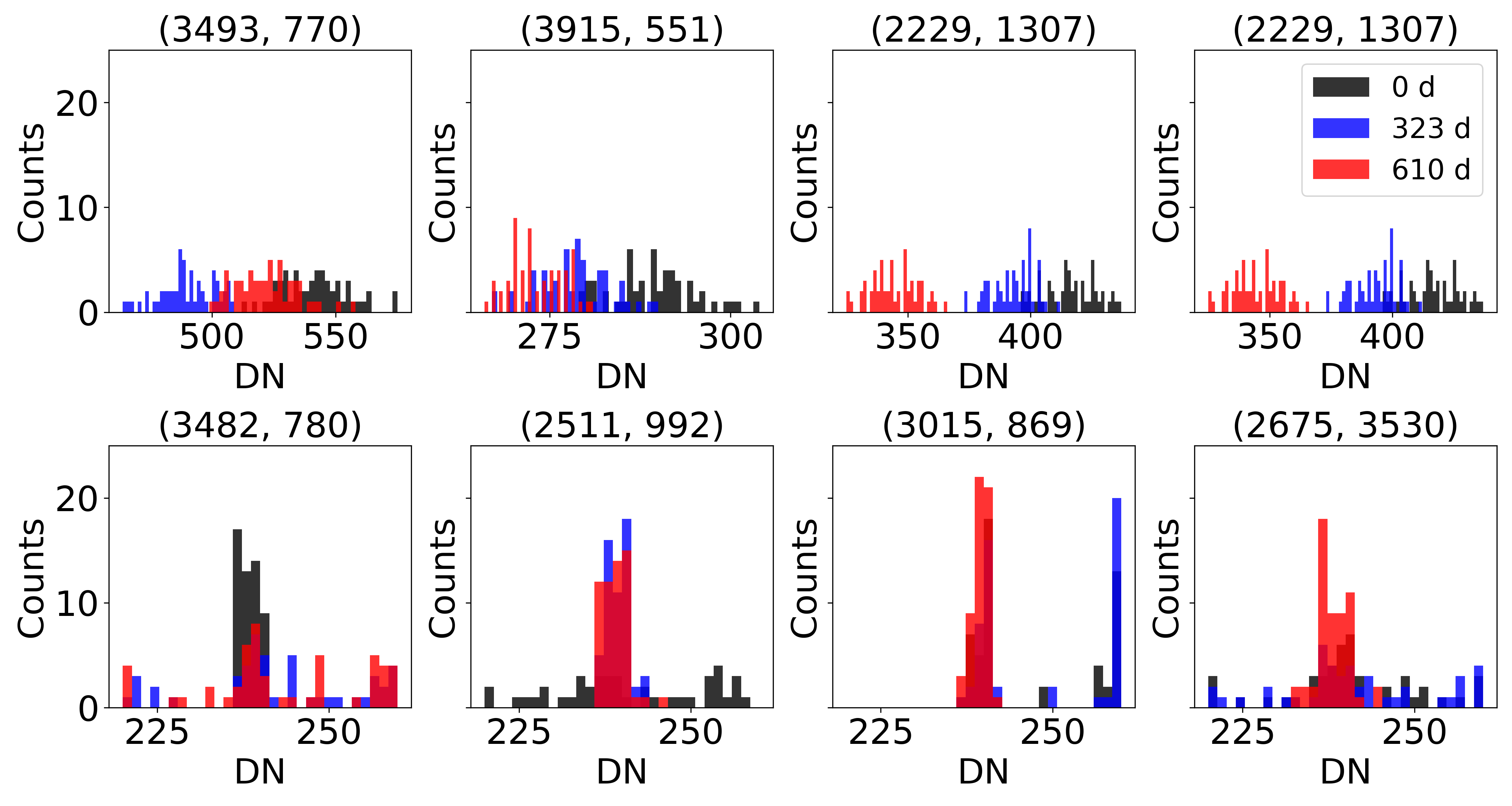}
    \caption{The 50-ms-bias distribution of eight degraded pixels at -30 $^\circ$C on the 0th day (black), the 323rd day (blue), and the 610th day (red) of the aging test, for bias-varied pixels (upper four panels) and noise-varied pixels (lower four panels), respectively. }
    \label{fig:pixel}
\end{figure*}

We randomly select eight degraded pixels, four for bias-varied and four for noise-varied, on the 610th day. We extract the DN values in each frame of these pixels. Their 50-ms-bias distributions and positions (Row, Column) are given in Figure \ref{fig:pixel}. For bias-varied pixels, it seems that there are two modes of bias variation. Some of them, like (2229, 1307), have continuously decreasing bias indicating the reduction of dark current by aging, while the others, like (3493, 770), show stochastic variation in bias during the aging process.  Some of the noise-varied pixels show multi-peak distributions, i.e., the random telegraph, at the beginning of the test. The random telegraph in these pixels, like (2511, 992) and (3015, 869), seems to be suppressed on the 610th day.

\section{Discussion and Conclusion}\label{sec:con}

CMOS sensors have been supposed to be the promising devices for X-ray astronomy. The long-term reliability of CMOS sensor is crucial for its application in space missions. In this work, we investigate the characteristics of a EP4K scientific CMOS sensor during a aging test for over 600 days. After aging at -30 $^\circ$C for 16 months then at 20 $^\circ$C for 5 months, the bias, dark current, readout noise, gain, and energy resolution of the sensor are almost unchanged. Dozens degraded pixels are found among the whole 4 k $\times$ 4 k pixels, which show a bias variation of over 10 DN or a noise change of over 15 e$^-$. Most of them are pixels with a decreased bias and the number of these pixels grows with the aging duration. Analysis show that aging process may have help to suppress the dark current and random telegraph signal of these pixels. In this study, the negligible impact of long-term aging effectively eliminates the contribution from the inherent decay of stability to the observed mild degradation under proton and $\gamma$-ray radiation \cite{2023JATIS...9d6003L,10.1117/1.JATIS.10.2.026001}. This work further emphasizes that possible variations in the in-orbit performances of scientific CMOS sensors may not be solely attributed to this negligible aging effect.

These performances firmly confirm the reliability of this CMOS sensor for a continue operation of several hundred days. With these test results, the lifetime of the sensor can be roughly estimated. The bias, dark current, and noise of the whole sensor are represented by the median values of the 4 k $\times$ 4 k pixels, with extremely small errors. Therefore, their weak variations in Figure \ref{fig:bias}, Figure \ref{fig:dc} and Figure \ref{fig:std} should not come from a long-term effect introduced by aging. Therefore, we focus on the evolution of the conversion gains and fit them via a first-order kinetic model \cite{WOS:000238158700003}, $G=G_0\exp{(-kt)}$, where $G$ is the value of gain at the time $t$, $G_0$ is the initial value of gain, and $k$ is the reaction kinetic constant. Using the -30 $^\circ$C gain values during the aging from February 2022 to June 2023, we obtain $k=2.4\times10^{-3}$ yr$^{-1}$, corresponding to a gain degeneration of 0.73\% over 3 years and 2.41\% over 10 years, which is low enough for most current and upcoming astronomical missions. This degeneration can be easily monitored and corrected by the annual calibration of gain for most X-ray telescopes.

\section*{Code and Data Availability}

The code and data underlying this article will be shared on reasonable request to the corresponding author.

\section*{Acknowledgments}

This work was supported by the National Natural Science Foundation of China (Grant Nos. 12173055 and 12333007) and the Chinese Academy of Sciences (Grant Nos. XDA15310000, XDA15310100, XDA15310300, and XDA15052100).




\end{document}